\documentclass{aip-cp}

\usepackage[numbers]{natbib}
\usepackage{rotating}
\usepackage{graphicx}

\begin{document}

% Title portion
\title{Symmetries of Neutrino Physics}

\author[aff1]{A.B. Balantekin\corref{cor1}}
\eaddress[url]{http://nucth.physics.wisc.edu/}

\affil[aff1]{University of Wisconsin, Department of Physics, Madison, Wisconsin 53706 USA}
\corresp[cor1]{Corresponding author: baha@physics.wisc.edu}

\maketitle

\begin{abstract}
In this contribution to the proceedings of a conference honoring the career of Francesco Iachello,  two applications of symmetry principles to neutrino physics are described. 
These applications are the connection between fermion pairing in many-body physics and the neutrino mass as well as collective neutrino oscillations. \end{abstract}

% Head 1

\section{INTRODUCTION}

Symmetry principles and associated algebraic techniques play a very special role in physics. Presence of symmetries determine if a problem is exactly solvable. Exactly solvable problems  are certainly interesting on their own. They may also be useful as toy models or may serve as a starting point of systematic approximations dealing with more complicated problems. Even breaking of symmetries leads to very interesting physics. In this contribution the utility of symmetry arguments is illustrated using two examples from neutrino physics: 
Symmetries of neutrino mass and collective neutrino oscillations. The natural mathematical framework to discuss symmetries is the language of Lie groups and Lie algebras. However, one does not necessarily need to employ complicated mathematical structures. In the examples chosen here it is shown that most of the salient physics can be recovered using the simplest Lie algebra, that of the angular momentum.

\section{NEUTRINO MASS AS A PAIRING PROBLEM}

A Dirac spinor with four independent components represents a Dirac fermion. The contribution of the mass of a Dirac fermion to the Lagrangian is given by 
\begin{equation}
{\cal L}_D = -m_D \left( \overline{\psi}_L \psi_R + \overline{\psi}_R  \psi_L \right)
\end{equation}
 where $m_D$ is the mass of the Dirac fermion and $\psi_L, \psi_R$ are the left- and right-handed components of the associated spinors.  In contrast, a  spinor describing a Majorana fermion is self charge-conjugate. Hence it is possible to assign different masses to the right-handed and left-handed components, resulting in the mass Lagrangian 
 \begin{equation}
 \label{majo}
{\cal L}_M = -\frac{m_L}{2} \left( \overline{\psi}_L \psi_L^C + {\rm h.c.} \right) -\frac{m_R}{2} \left( \overline{\psi}_R \psi_R^C + {\rm h.c.} \right) 
\end{equation}
where the spinor $\psi^C$ is the charge-conjugate of the spinor $\psi$ and $m_{L,R}$ are the two independent Majorana masses. The factor $1/2$ is introduced to avoid double counting. 
Since the mass terms in Equation (\ref{majo}) violate charge conservation, Majorana masses are only possible for neutral fermions, such as neutrinos. In the Standard Model of particle physics all masses of the massive elementary particles, possibly except those for neutrinos, are Dirac masses generated by the Yukawa couplings of the Higgs. A recent review surveying various properties of neutrinos, including their masses, is given in Reference \cite{Balantekin:2018azf}. 

Hamiltonians associated with the mass Lagrangians given above are reminiscent of the pairing Hamiltonians in nuclear and condensed-matter physics. Indeed the analogy between 
fermion mass and pairing gap in the theory of superconductivity was already discussed a long time ago in exploring consequences of spontaneous symmetry breaking. \cite{Nambu:1961fr,Nambu:1961tp}
Pairing is ubiquitous in many-fermion systems. For example in nuclear physics s-wave ($L=0$) pairing of nucleons in a single Shell Model orbital $j$ is described by the quasi-spin algebra:
\begin{eqnarray}
\hat{S}^+_j&=&\sum_{m>0} (-1)^{(j-m)} a^\dagger_{j\>m}a^\dagger_{j\>-m},
\nonumber \\
\hat{S}^-_j&=&\sum_{m>0} (-1)^{(j-m)} a_{j\>-m}a_{j\>m}, \nonumber
\end{eqnarray}
\begin{equation}
\hat{S}^0_j=\frac{1}{2}\sum_{m>0}
\left(a^\dagger_{j\>m}a_{j\>m}+a^\dagger_{j\>-m}a_{j\>-m}-1
\right) .
\label{quasispin}
\end{equation}
where $m$ takes values $-j \le m \le +j$.  This algebra is an SU(2) algebra the quadratic Casimir operator of which is  $\Omega_j (\Omega_j/2+1)/2$, where $\Omega_j = j+1/2$. 
Once can also introduce d-wave ($L=2$) pairing, enlarging the algebra to SU(6) and leading to the celebrated Interacting Boson Model describing the structure of  low-lying states of medium-heavy nuclei  \cite{Arima:1981hp}. 

To establish the connection with the fermion masses we introduce massless fermion fields 
\begin{equation}
\hspace*{-2.5cm}
\psi_L ( \mathbf{x}, t) = \int    \frac{d^3\mathbf{p}}{2E (2\pi)^3}  \left[ a(\mathbf{p} , h=-1) \> u(\mathbf{p} , h=-1) e^{-ip.x} + b^{\dagger} (\mathbf{p} , h=+1) \> v( \mathbf{p} , h=+1) e^{ip.x} \right] ,
\end{equation}
and
\begin{equation}
\hspace*{-2.5cm}
\psi_R ( \mathbf{x}, t) = \int \frac{d^3\mathbf{p}}{2E (2\pi)^3}   \left[ a(\mathbf{p} , h=+1) \> u(\mathbf{p} , h=+1) e^{-ip.x} + b^{\dagger} (\mathbf{p} , h=-1) \>  v( \mathbf{p} , h=-1) e^{ip.x} \right] ,
\end{equation}
where $h$ is the helicity, $u$ and $v$ are the Dirac spinors in the helicity basis, $a$ and $b$ are the associated creation-annihilation operators. We then can write down the mass term of the free fermion Hamiltonian as 
\begin{equation}
\label{massterm}
m_D  \int d^3x\> \left( \overline{\psi}_L \psi_R +  \overline{\psi}_R \psi_L \right) = m_D \sum_h \int \frac{d^3\mathbf{p}}{2E (2\pi)^3}  \left[ a^{\dagger} (\mathbf{p} , h) b^{\dagger} (- \mathbf{p} , h) +
b( - \mathbf{p} , h)a (\mathbf{p} , h) \right] .
\end{equation}
One observes the similarity between Dirac mass term in Equation (\ref{massterm}) and the quasispin algebra of Equation (\ref{quasispin}). In fact such a mass term can be written as the sum of the ladder operators of an SU(2) algebra, which we designate as SU(2)$_D$. One can of course also introduce such SU(2) algebras for the Majorana mass terms made out of left- and right-handed fermion fields: SU(2)$_L$ and SU(2)$_R$. Since the left- and right-handed components of a fermion field are independent,  SU(2)$_L$ and SU(2)$_R$ algebras commute with each other (but of course not with SU(2)$_D$). In general from four independent components of a Dirac spinor one expects to form an Sp(4) $\sim$ SO(5) algebra 
\cite{Balantekin:2000qt}. The SU(2)$_L \times$ SU(2)$_R \sim$ SO(4)$_{LR}$  is the maximum subalgebra of this SO(5) algebra. It is known that the SO(5) algebra includes four different SU(2) algebras, not all of which are mutually commuting. Three of these subalgebras are SU(2)$_L$, SU(2)$_R$, and SU(2)$_D$. It turns out that this fourth SU(2) algebra, 
labeled here as SU(2)$_{\rm PG}$, generates the 
Pauli-G\"ursey transformation \cite{PG}
\begin{equation}
\psi\rightarrow\psi^{\prime} = \hat{U} \psi \hat{U}^{\dagger} =a\psi+b\gamma_5\psi^c,  \>\> |a|^2+|b|^2=1 
\end{equation} 
where $\hat{U}$ is the SU(2)$_{\rm PG}$ group rotation associated with this fourth SU(2) subalgebra of SO(5) \cite{Balantekin:2000qt}.

\section{COLLECTIVE NEUTRINO OSCILLATIONS}

It is a well-established experimental fact that the neutrino flavor eigenstates, which participate in the weak interactions, are linear combinations of mass eigenstates. As neutrinos travel in vacuum, this mixing leads to interference between two or more quantum states, which is observed as flavor oscillations. In the presence of background particles with relatively low densities  one usually ignores collisions of neutrinos with these background particles as the cross sections are very small (proportional to the square of the Fermi constant, G$_F^2$). However the {\em amplitude} of the coherent forward scattering of neutrinos off the background particles, proportional to G$_F$, could contribute to the phases of the neutrino wavefunctions which interfere with each other. The resulting phenomena is referred to as the matter-enhanced neutrino oscillations.  

For two flavors of neutrinos, labeled as e and x, one can introduce creation and annihilation operators for a neutrino 
with three momentum ${\bf p}$, and write down the generators of an SU(2) algebra representing neutrino flavor isospin 
\cite{Balantekin:2006tg}: 
\begin{eqnarray}
J_+({\bf p}) &=& a_x^\dagger({\bf p}) a_e({\bf p}), \> \> \>
J_-({\bf p})=a_e^\dagger({\bf p}) a_x({\bf p}), \nonumber \\
J_0({\bf p}) &=& \frac{1}{2}\left(a_x^\dagger({\bf p})a_x({\bf p})-a_e^\dagger({\bf p})a_e({\bf p})
\right). \label{su2}
\end{eqnarray}
Using the operators in Equation (\ref{su2}) 
the Hamiltonian for neutrinos traveling through static, locally charge-neutral, and unpolarized matter takes the form  
\begin{equation}
\label{msw}
 H_{\nu} = \int d^3{\bf p} \frac{\delta m^2}{2p} \left[
\cos{2\theta} \> J_0({\bf p}) + \frac{1}{2} \sin{2\theta}
\left(J_+({\bf p})+J_-({\bf p})\right) \right] -  \sqrt{2} G_F \int d^3{\bf p} 
\> N_e \>  J_0({\bf p}) 
\end{equation}
where $\theta$ is the mixing angle between two neutrino flavor states, and $N_e$ is the electron density of the background. Defining $\omega_p = \delta m^2/ 2p$ and introducing an  auxiliary vector $\mathbf{B}$ in the flavor basis 
\begin{equation}
{\mathbf B}_{\rm flavor} = (\sin 2 \theta,0, -\cos 2 \theta). 
\end{equation}
the Hamiltonian in Equation (\ref{msw}) takes the form
\begin{equation}
\label{msw}
 H_{\nu} = \int d^3{\bf p} \> \left( \omega_p \>  \mathbf{B} \cdot \mathbf{J} (\mathbf{p}) -  \sqrt{2} G_F  
\> N_e \>  J_0({\bf p}) \right) .
\end{equation}
Note that the auxiliary vector  $\mathbf{B}$ takes the form 
\begin{equation}
{\mathbf B}_{\rm mass} = (0,0, -1)
\end{equation}
in the mass basis. 

In many astrophysical sites, such as core-collapse supernovae and merging neutron stars, an abundant number of neutrinos are present. For example the entire gravitational binding energy of a pre-supernova star is deposited into the proto-neutron star and is converted into neutrinos, yielding $\sim 10^{57}-10^{58}$ neutrinos. In such situations one can no longer ignore coherent forward scattering of neutrinos off other neutrinos and one has to add a second term to the Hamiltonian in Equation (\ref{msw}) to account for this effect:
\begin{equation}
\label{nunu}
H_{\nu \nu} = \sqrt{2} \frac{G_F}{V} \int d^3{\bf p} \> d^3{\bf q} \>  (1-\cos\vartheta_{\bf pq}) \> {\bf
J}({\bf p}) \cdot {\bf J}({\bf q}) ,
\end{equation}
where $\vartheta_{\bf pq}$ is the angle between neutrino momenta {\bf p} and {\bf q} and V 
is the normalization volume.  Note that the Hamiltonian in Equation (\ref{msw}) includes a ``kinetic energy" term and a one-body interaction of the neutrino with the electron background. 
In contrast the Hamiltonian in Equation (\ref{nunu}) includes a genuine two-body interaction. 

To calculate propagation of a very large number of neutrinos subject to the Hamiltonian $H_{\nu} + H_{\nu \nu}$ is indeed a formidable task and typically many approximations are employed. Near the proto-neutron star in a core-collapse supernova, the sheer number of neutrinos overwhelms the interactions of neutrinos with background electrons; one can then 
ignore $N_e$ term and write
\begin{equation}
\label{totalexact}
H = \int d^3{\bf p} \>  \omega_p \>  \mathbf{B} \cdot \mathbf{J} (\mathbf{p}) + \frac{\sqrt{2} \>G_F}{V} \int d^3{\bf p} \> d^3{\bf q} \>  (1-\cos\vartheta_{\bf pq}) \> {\bf
J}({\bf p}) \cdot {\bf J}({\bf q}) . 
\end{equation}
(Of course in the outer shells of the supernova, the neutrino flux is significantly diminished. In that region the term with $N_e$ is significantly larger than the $H_{\nu \nu}$ term and cannot be ignored). It is straightforward to include a third flavor of neutrinos (by enlarging the SU(2) algebra to SU(3) algebra) and introduce antineutrinos  (by introducing a second set of SU(3) algebras to describe antineutrinos). 

A commonly utilized approximation is the mean field approximation, replacing products of operators with one single operator and the expectation value of the second operator. This approximation can be rigorously derived from the path-integral formalism as the saddle-point approximation \cite{Balantekin:2006tg}. Recalling the neutrino-neutrino interaction is a current-current interaction, the interaction of Fierz-transformed neutrino currents is approximated as 
\begin{equation}
\overline{\psi}_{\nu L} \gamma^{\mu} \psi_{\nu L} \overline{\psi}_{\nu L} \gamma_{\mu} \psi_{\nu L} \rightarrow \overline{\psi}_{\nu L} \gamma^{\mu} \psi_{\nu L} \langle \overline{\psi}_{\nu L} \gamma_{\mu} \psi_{\nu L} \rangle + \cdots
\end{equation}
where the averages are calculated using the SU(2) coherent states associated with the algebra in Equation (\ref{su2}) \cite{Balantekin:2006tg}. The SU(2) coherent state path-integral formalism in the saddle-point approximation gives the interaction of Fierz-transformed neutrino-antineutrino currents as 
\begin{equation}
\overline{\psi}_{\nu L} \gamma^{\mu} \psi_{\nu L} \overline{\psi}_{\nu R} \gamma_{\mu} \psi_{\nu R} \rightarrow \overline{\psi}_{\nu L} \gamma^{\mu} \psi_{\nu L} \langle \overline{\psi}_{\nu R} \gamma_{\mu} \psi_{\nu R} \rangle + \cdots .
\end{equation}
It was pointed out that neutrino-antineutrino interaction can give rise to another mean field \cite{Vaananen:2013qja,Serreau:2014cfa,Vlasenko:2013fja,Cirigliano:2014aoa}: 
\begin{equation}
\overline{\psi}_{\nu L} \gamma^{\mu} \psi_{\nu L} \overline{\psi}_{\nu R} \gamma_{\mu} \psi_{\nu R} \rightarrow \overline{\psi}_{\nu L} \gamma^{\mu} \langle \psi_{\nu L}  \overline{\psi}_{\nu R} \gamma_{\mu} \rangle  \psi_{\nu R}  + \cdots .
\end{equation}
Noting that $\langle \psi_{\nu L}  \overline{\psi}_{\nu R} \rangle \propto m_{\nu}$  from the symmetry principles, one concludes that effects of such a mean field would be negligible if the medium is isotropic \cite{Tian:2016hec}. 

Many calculations in the mean-field approximation are available in the literature (for a review see Reference \cite{Duan:2010bg}). These calculations observed that at a particular energy the final neutrino spectra emerging from the collective neutrino oscillations are almost completely divided into parts of different flavors. This observations was termed  spectral swappings or splits  \cite{Raffelt:2007cb,Duan:2008za}. Spectral swaps and other effects resulting from collective neutrino oscillations can have significant impact on astrophysics. For example, a recent mean-field calculation of collective neutrino oscillations with three flavors connected with a nucleosynthesis network calculation explored $\nu$p process nucleosynthesis in proton-rich neutrino-driven winds. It found that there could be a significant enlargement of abundances of p-nuclei \cite{Sasaki:2017jry}. 

For a Hamiltonian describing $N$ particles each of which can occupy $p$ states the dimension of the Hilbert space is $p^N$. Exact solutions for the eigenstates of such a Hamiltonian 
include both entangled and unentangled states. Adopting a mean-field approximation eliminates the entangled states and reduces the dimension of the Hilbert space to $pN$. Hence, exact solutions are desirable for a complete description of physics even in the limiting cases. One such limiting case is the single-angle approximation. In this approximation the quantity $(1-\cos\vartheta_{\bf pq})$ is replaced by its average value over the ensemble:
\begin{equation}
\frac{\sqrt{2} \>G_F}{V}  (1-\cos\vartheta_{\bf pq}) \rightarrow \frac{\sqrt{2} \>G_F}{V} \langle (1-\cos\vartheta_{\bf pq}) \rangle \equiv \mu 
\end{equation}
leading to the Hamiltonian
\begin{equation}
\label{singleangle}
H = \sum_p \omega_p \mathbf{B} \cdot \mathbf{J} (\mathbf{p}) +  \mu \sum_{p,q,p \neq q}  {\bf
J}({\bf p}) \cdot {\bf J}({\bf q})  .
\end{equation} 
The Hamiltonian in Equation (\ref{singleangle}) is exactly solvable \cite{Pehlivan:2011hp} using the Bethe ansatz method as we describe in the next section.

\section{GAUDIN ALGEBRA AND BETHE ANSATZ}

To apply the Bethe ansatz method it is more convenient to consider the discretized version in  Equation (\ref{singleangle}).  
Here we summarize only the salient points of the application of Bethe ansatz method to the problem of collective neutrino oscillations. For a more detailed summary, the reader is referred to a recent review \cite{Balantekin:2018mpq}. 
In general the Gaudin algebra is defined by the commutation relations \cite{gaudin}
\begin{equation}
\label{c1}
[S^+(\lambda),S^-(\rho)]=2\frac{S^0(\lambda)-S^0(\rho)}{\lambda-\rho},
\end{equation}
\begin{equation}
\label{c2}
[S^0(\lambda),S^{\pm}(\rho)]=\pm\frac{S^{\pm}(\lambda)-S^{\pm}(\rho)}
                                                  {\lambda-\rho},
\end{equation}
\begin{equation}
\label{c3}
[S^0(\lambda),S^0(\rho)]=[S^{\pm}(\lambda),S^{\pm}(\rho)]=0 .
\end{equation}
In the above equations $\lambda$ and $\rho$ are arbitrary parameters which can take both real and complex values. For the neutrino physics the appropriate realization of this algebra is given by 
\begin{equation}
\label{gaudinwithsu2}
S^{0}(\lambda)= \frac{1}{\mu} + \frac{1}{2}\sum_{p}\frac{ a_x^\dagger({\bf p})a_x({\bf p})-a_e^\dagger({\bf p})a_e({\bf p})}{\omega_p-\lambda} 
\end{equation}
\begin{equation}
\label{gaudinwithsu22}
S^+(\lambda)=\sum_{p} \frac{ a_x^\dagger({\bf p}) a_e({\bf p})}{\omega_p-\lambda} , \>\>\> \>\>\> S^-(\lambda)=\sum_{p} \frac{ a_e^\dagger({\bf p}) a_x({\bf p})}{\omega_p-\lambda} . 
\end{equation}
One can show that the operators 
\begin{equation}
\label{c4}
X(\lambda)=S^0(\lambda)S^0(\lambda)+\frac{1}{2}S^+(\lambda)S^-(\lambda)+
           \frac{1}{2}S^-(\lambda)S^+(\lambda)
\end{equation}
commute for different values of the parameters:
\begin{equation}
\label{commute}
[X(\lambda),X(\rho)]=0  , \>\>\> \lambda \neq \rho. 
\end{equation} 

Using the realization given in Equations (\ref{gaudinwithsu2}) and (\ref{gaudinwithsu22}) we can write 
\begin{equation}
\label{b6}
X(\lambda) = \sum_p \frac{\mathbf{J}^2 (\mathbf{p})}{(\omega_p - \lambda)^2} + {\cal H} (\lambda) + \frac{1}{\mu^2}
\end{equation}
where we defined
\begin{equation}
\label{b8}
{\cal H}(\lambda) = \sum_{p,q,p\neq q}  \frac{\mathbf{J} (\mathbf{p}) \cdot \mathbf{J} (\mathbf{q})}{( \omega_p-\lambda)(\omega_q-\lambda)} + \frac{2}{\mu} \sum_p \frac{J^0(\mathbf{p})}{(\omega_p-\lambda)} . 
\end{equation}
Taking the limit 
\begin{equation}
\label{b9}
\lim_{\lambda \rightarrow \omega_p} (\lambda - \omega_p) {\cal H}(\lambda) = 2 \sum_{q,q\neq p} \frac{\mathbf{J} (\mathbf{p}) \cdot \mathbf{J} (\mathbf{q})}{\omega_p - \omega_q} - \frac{2}{\mu} {J}^0 (\mathbf{p})  
\end{equation}
and recalling Equation (\ref{commute}) we get the following mutually commuting operators, one for each $p$:  
\begin{equation}
\label{b12}
\frac{2h_p}{\mu} = 2 \sum_{q,q\neq p} \frac{\mathbf{J} (\mathbf{p}) \cdot \mathbf{J} (\mathbf{q})}{\omega_p - \omega_q} - \frac{2}{\mu}  J^0 (\mathbf{p}) . 
\end{equation}
Multiplying Eq. (\ref{b12}) with $\omega_P$ and summing over $p$ in Eq. (\ref{b12}) gives the Hamiltonian of Eq. (\ref{singleangle}) written in the mass basis:
\begin{equation}
\label{b13}
\frac{H}{\mu} = \sum_p \omega_p \frac{h_p}{\mu} =  \sum_{q,p, q\neq p} \mathbf{J} (\mathbf{p}) \cdot \mathbf{J} (\mathbf{q}) 
- \frac{1}{\mu} \sum_p \omega_p J^0 (\mathbf{p}) . 
\end{equation}

Eigenstates of the Hamiltonian in Equation (\ref{singleangle}) are given by 
\begin{equation}
\label{unnorstate}
|\xi> \equiv
|\xi_1,\xi_2,\dots,\xi_n> \propto  S^+(\xi_1) S^+(\xi_2)\dots
S^+(\xi_n)|0>
\end{equation}
if the complex numbers
$\xi_1,\xi_2,\dots, \xi_n$ satisfy the Bethe 
Ansatz equations:
\begin{equation}
\label{Ba}
\frac{1}{\mu} - \sum_p \frac{j_p}{\omega_p - \xi_{\alpha}} = \sum_{ {\beta=1}\atop{(\beta\neq\alpha)} }^n
\frac{1}{\xi_\alpha-\xi_\beta} \quad \mbox{for} \quad
\alpha=1,2,\dots,n.
\end{equation} 
In Equation (\ref{unnorstate}) is the state annihilated by $S^- (\lambda)$ for all values of $\lambda$. In Equation (\ref{Ba}), $j_p$ is the SU(2) label for the algebra spanned by $\mathbf{J} (\mathbf{p})$ with the Casimir operator $j_p (j_p+1)$. 
The eigenvalues of the Hamiltonian in Equation (\ref{singleangle}) are given by 
\begin{equation}
E_n = \mu \sum_{p \neq q} j_pj_q + \frac{1}{2} \sum_p \omega_p j_p - \mu n \sum_p j_p + \mu \frac{n(n-1)}{2} - \frac{1}{2} \sum_{\alpha} \xi_{\alpha}. 
\end{equation}
It is again possible to extend the formalism presented above to include antineutrinos and three-flavor mixing including CP-violating phases \cite{Pehlivan:2014zua}. 

One reason to seek an exact solution to the problem is to test the limits of validity of the mean-field approximations. A recent work explored adiabatic evolution of an initial thermal distribution of electron neutrinos (T=10 MeV) and antineutrinos of another flavor (T=12 MeV) using the exact formalism described in this section. It considered 108 neutrinos and antineutrinos distributed over 1200 energy bins both for neutrinos and antineutrinos with solar neutrino parameters and normal hierarchy. It found that spectral split phenomenon is still present even in the exact adiabatic calculation \cite{Birol:2018qhx}. 

\subsection{Solving Bethe ansatz equations}

Many approaches are proposed to solve the Bethe ansatz equations. Here we outline a technique which goes back to Stieltjes \cite{still}. In this approach one starts writing down a polynomial the roots of which are the solutions of the Bethe ansatz equations, $\xi_{\alpha}$: 
\begin{equation}
\label{plam}
P(\lambda) = \prod_{\alpha} (\lambda- \xi_{\alpha}) = \exp \left( \sum_{\alpha} \log (\lambda -\xi_{\alpha}) \right) .
\end{equation}
Next one introduces
\begin{equation}
\Lambda (\lambda) = \frac{dP/d\lambda}{P(\lambda)} = \sum_{\alpha=1}^N \frac{1}{\lambda - \xi_{\alpha}} .
\end{equation}
Assuming that the variables in Equation (\ref{plam}) are solutions of the Bethe ansatz equation in Equation (\ref{Ba}), one can 
show that the quantity $\Lambda (\lambda)$ satisfies the following differential equation: 
\begin{equation}
\label{mastereq}
\Lambda^2 (\lambda) + \frac{d\Lambda}{d\lambda} + \frac{1}{\mu} \Lambda (\lambda)= 2 \sum_p  \frac{j_p}{\lambda - \omega_p} \left[ \Lambda(\lambda) - \Lambda (\omega_p) \right] .
\end{equation}
This is a non-linear equation, reflecting the fact that the collective neutrino oscillations it derives from is a non-linear problem. An idea which was pursued in the literature by many authors is 
to calculate $\Lambda (\lambda) $ and its derivatives with respect to $\lambda$ only for $\lambda = \omega_p$ (see for example Reference \cite{Babelon:2007td}). 
Such a substitution yields the equation
\begin{equation}
\Lambda^2 (\omega_q) + (1-2 j_q) \Lambda' (\omega_q) + \frac{1}{\mu} \Lambda (\omega_q)= 2 \sum_{p \neq q}  j_p 
\frac{\Lambda(\omega_q ) - \Lambda (\omega_p)}{\omega_q - \omega_p} ,
\end{equation}
where prime denotes derivative with respect to $\lambda$. For $j_q=1/2$ the derivative term vanishes yielding an algebraic equation for $\Lambda (\omega_q) $.  If all the $j_p$ are also $1/2$, then one has a set of algebraic equations for all $\Lambda (\omega_p)$. For higher values of $j_p$ if one keeps taking derivatives, one eventually reaches to vanishing terms. For example, taking the next derivative produces a second derivative term which vanishes for $j_q=1$: 
\begin{equation}
2 \Lambda (\omega_q) \Lambda'(\omega_q) + (1-j_q)\Lambda''(\omega_q) + \frac{\Lambda'(\omega_q)}{\mu}  
= 2 
\sum_{p \neq q} j_p \left[ \frac{\Lambda'(\omega_q)}{\omega_q - \omega_p} + \frac{\Lambda(\omega_q) - \Lambda(\omega_p)}{(\omega_q - \omega_p)^2} \right] .
\end{equation}
An application of this approach to adiabatic collective neutrino oscillations is in progress and will be presented elsewhere \cite{AMA}. 

\section{CONCLUSION}

Symmetries play a key role in understanding many physical phenomena as aptly illustrated by the body of scientific work of Franco Iachello. In this contribution two examples of the applications of the symmetry principles to neutrino physics are presented. Both examples, connection between neutrino mass and pairing problem as well as the collective neutrino oscillations, are chosen from active research in neutrino physics and astrophysics.

% Sections that will go in second font

% Acknowledgement
\section{ACKNOWLEDGMENTS}
The author would like to thank Franco Iachello for his mentorship and for many physics discussions over four decades covering all aspects of symmetries. 
This work was supported in part by the US National Science Foundation (NSF) Grants No. PHY-1806368 and PHY-1630782. 
A Franco, molte grazie e tanti migliori auguri per il prossimo capitolo della tua vita!

% References

\nocite{*}
\bibliographystyle{aipnum-cp}%
\bibliography{sample}%

\end{document}